# Association of Mid-Infrared Solar Plages With Calcium K Line Emissions and Magnetic Structures

R. Marcon [1]
Institute of Physics "Gleb Wataghin", State University of Campinas, Campinas, Brazil

P. Kaufmann[2,3], A. M. Melo [2], A. S. Kudaka
Center of Radio Astronomy and Astrophysics, Mackenzie Presbyterian University, São Paulo, Brazil

E. Tandberg-Hanssen
Center for Space Plasma and Aeronomic Research, University of Alabama in Huntsville, AL, USA

**ABSTRACT.** . Solar mid-IR observations in the 8-15 μm band continuum with moderate angular resolution (18 arcseconds) reveal the presence of bright structures surrounding sunspots. These plage-like features present good association with calcium CaII K1v plages and active region magnetograms. We describe a new optical setup with reflecting mirrors to produce solar images on the focal plane array of uncooled bolometers of a commercial camera preceded by germanium optics. First observations of a sunspot on September 11, 2006 show a mid-IR continuum plage exhibiting spatial distribution closely associated with CaII K1v line plage and magnetogram structures. The mid-IR continuum bright plage is about 140 K hotter than the neighboring photospheric regions, consistent with hot plasma confined by the magnetic spatial structures in and above the active region.

## 1. INTRODUCTION

The first solar observations in the mid-IR continuum (8-12 μm) were obtained using large telescopes with high angular resolutions (few arcseconds, Turon & Léna 1970; Gezari, Livingstone & Varosi 1999). These observations measure plasma temperatures above the photosphere, presenting advantages over lines in the visible because they are free from fluctuations caused by micro-turbulences and from Doppler enlargements of spectral lines. Mid-IR continuum images can become an important complementary observational tool to describe spatial features of hot plasma above the photosphere in active regions. Bright mid- IR area surrounding a sunspot was recently reported, from observations made with a small- aperture telescope and wide photometric beam (25 arcseconds) (Melo et al. 2006). They were suggested as equivalent to the previously found "plage-like" features (Gezari, Livingstone & Varosi 1999).

Temperatures in the low chromosphere are not possible to be measured directly. They may be theoretically inferred approximately from the analysis of the CaII K 3933.684 Å line (Linsky 1968). It shows wide absorption wings, with width of about 10 Å, and enhanced central intensity over higher temperature regions. It is suitable to describe temperature distributions over solar plages, usually associated with active regions and sunspots, where magnetic fields are more intense and plasma temperatures are higher. Thus, the CaII K1v spectral profile represents temperatures at altitudes ranging from 500 to 1000 km in the chromosphere (Vernazza, Avrett & Loeser 1981; Ulmschneider 2003). Accordingly to unpublished calculations by J. Jefferies and C. Lindsey (quoted by Gezari, Livingstone & Varosi 1999), the following approximated altitudes and temperatures might be assumed for corresponding mid-IR wavelengths: 70 km, 5700 K at 4.8 μm; 180 km, 5100 K at 12.4 μm and 250 km, 4800 K at 18.1 μm, respectively. Images of CaII K plages are also known to exhibit spatial structures closely related to the magnetic fluxes emerging from the photosphere (Babcock & Babcock 1955; Leighton 1959; Schrijver et al. 1989).

---

[1] "Bernard Lyot" Solar Observatory, Campinas, Brazil
[2] Center of Semiconductors Components, State University of Campinas, Campinas, Brazil
[3] e-address: pierre.kaufmann@pq.cnpq.br

On the other hand, the mid-infrared continuum emissions present intensities linearly proportional to the temperature. In this paper we will show the spatial correlation between the magnetic configuration, the CaII K1v line higher temperature plage and the mid-IR continuum images obtained with a new optical setup, using an uncooled microbolometer focal-plane array (FPA) camera centered at 10 μm.

## 2. THE OPTICAL SETUP FOR 10 μM SOLAR OBSERVATIONS

The optical arrangement assembled to perform the mid-IR solar observations is shown in Fig. 1. It consists of a 150 mm diameter parabolic reflector with 1200 mm primary focal length. The solar image is deflected by two flat mirrors into a rectangular 3 x 4 mm diaphragm opening in an aluminum plate acting as a heat dissipater. This concept is similar to the one used in the Dutch Open Telescope (Rutten et al. 2004). It is intended to prevent overheating of the IR camera optics. The setup is followed by a germanium lens to transfer the solar image to the focal-plane array of microbolometers of a FLIR A20 camera, containing 320 x 240 pixels (FLIR 2007). The camera`s own readout electronics and software feed the pixel readouts into a microcomputer. The photometric beam of the setup corresponds to the diffraction limit, being 18 arcseconds.

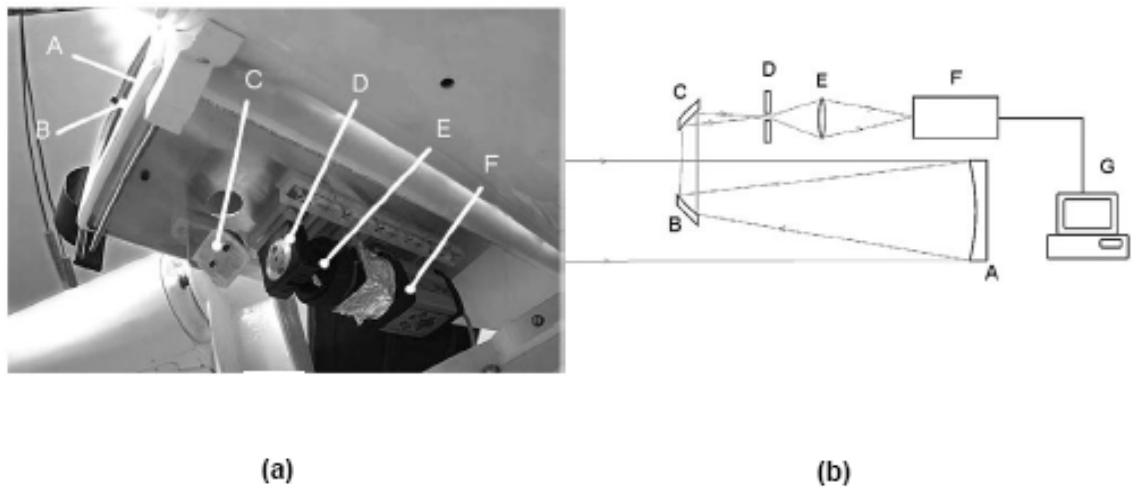

Fig. 1 – The 10 μm band instrumental setup, as assembled in (a) and in a simplified schematic diagram in (b). Labels A: primary parabolic mirror (150 mm in diameter, f/8); B and C: deflecting flat mirrors; D: thermal dissipater diaphragm; E: germanium lens; F: mid IR camera with uncooled microbolometer FPA and G: data acquisition microcomputer.

## 3. OBSERVATIONS AND DATA ANALYSIS

The first results using the above described instrumentation were obtained on September 11, 2006, with solar observations carried out on AR 904, at about 13:00 UT. The signal-to-noise ratio of the images was considerably improved using the technique of division by a flat-field frame obtained on a quiet area in the solar disk. The camera was operated in video mode, setting the minimum temperature level above the dark frame level, which contribution becomes negligible. Nearly 300 photograms were captured at a rate of 15 frames-per-second on the active region; another sequence was captured on a quiet region, as illustrated in Fig. 2. The latter has been averaged to become a comparison mean flat-field frame. Every frame on the active region movie was then divided by the flat-field, aligned, added to each other and averaged. The final result is shown in Fig. 3 (a). The mid-IR photometric beam is shown at the bottom right. The figure also shows in (b) a magnetogram, obtained by the National Solar Observatory (VSM- NSO/SOLIS 2006 ) and in (c) a CaII K1v image obtained at Observatoire de Paris-Meudon (BASS2000 2006). The similarities in morphology in the three images are striking.

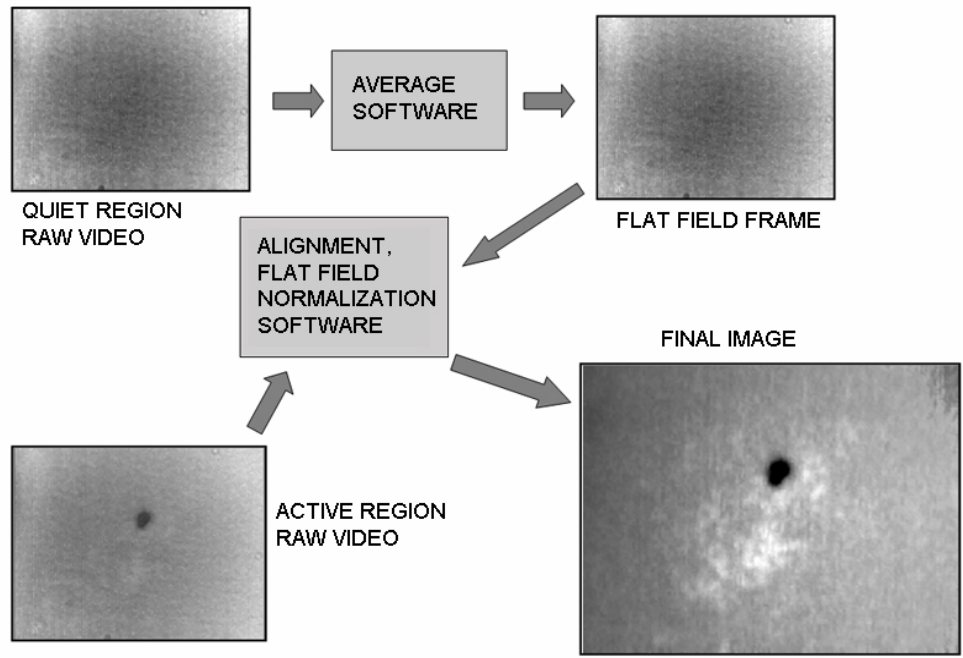

Fig. 2 – Procedures and steps to produce the flat-field on a quiet area in the solar disk, the active region frames alignment and division by the flat-field to obtain the final image with higher signal-to-noise ratio and resolution.

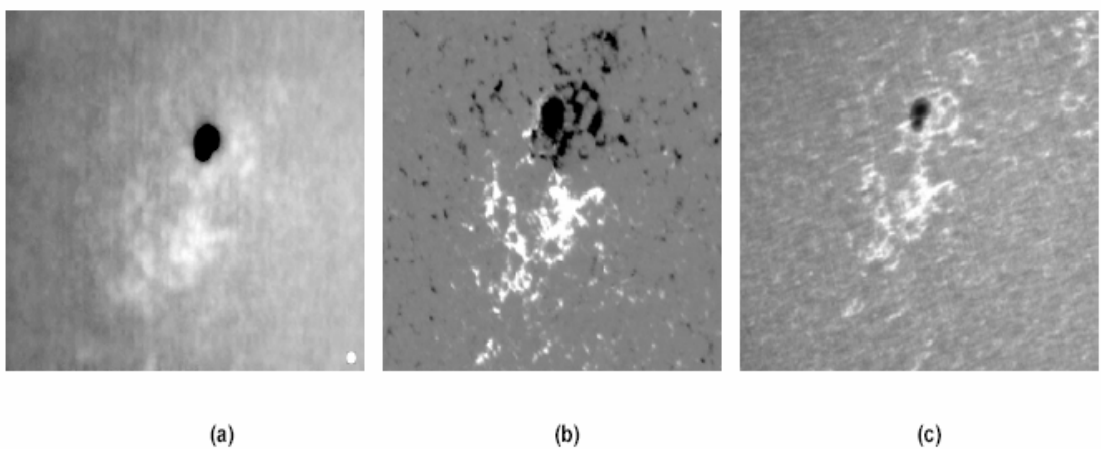

Fig. 3 – (a) 10.8 x 8.3 arcmin image of AR 904 on September 11, 2006 at about 13:00 UT, (b) a National Solar Observatory magnetogram (VSM- NSO/SOLIS 2006), (c) and an Observatoire de Paris-Meudon CaII K1v image (BASS2000 2006)   The size of the 10 μm photometric beam in shown in the bottom-right of image (a).

A photometric profile has been drawn on the active region, shown in Fig. 4 (a). The profile is plotted in Fig. 4 (b), scaling the photosphere-umbra for the 10 μm range after Turon & Lena (1970). The mid-IR plage temperature excess above the photosphere is well defined, being about 140 K above the photospheric values, in the adopted Turon & Léna (1970) scale.

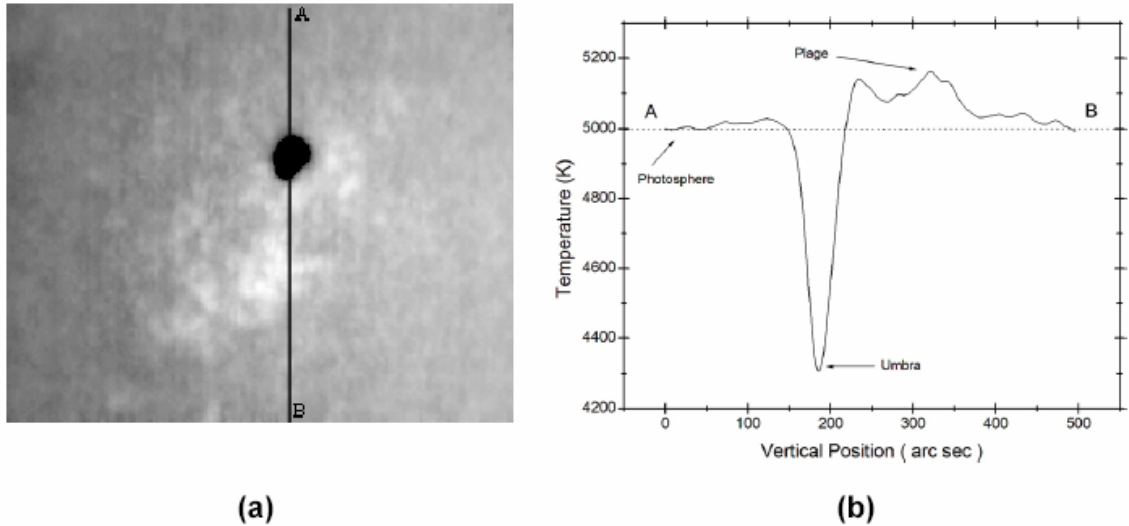

Fig. 4 – A photometric profile drawn on the 10 μm image (a), is plotted in (b). The temperature range was scaled after the photosphere-umbra temperatures obtained by Turon & Léna (1970) in the same wavelength range. The mid -IR plage is about 140 K hotter than the surrounding photosphere.

### 4. CONCLUDING REMARKS

It has been shown that photometry of solar active regions in the mid-IR continuum, obtained with relatively simple instrumentation providing moderate space resolution, is a powerful tool to describe quiescent plages and magnetic field. The presence of bright areas associated with a sunspot is confirmed, exhibiting striking similarity to CaII K and magnetic field morphologies. These observations present advantages compared to other techniques in the visible because they are free from problems caused by Doppler enlargement of spectral lines in monochromatic observations. As pointed out by Turon & Léna, the seeing at the 10 μm band is considerably better than in the visible. We suggest that regular mid-IR continuum solar observations might become a new promising tool for monitoring the development of quiescent solar active regions.

This research received partial support from Brazilian agencies FAPESP (Proc. nos.03/07746-6;04/07835-1), CNPq (471220/2004; 305034/2006-5; 372631/2006-1) and MackPesquisa (edital 2006). Authors gratefully acknowledge for the useful suggestions given by an anonymous referee.